\documentclass[10pt,twocolumn,superscriptaddress,english,10pt,prl,showpacs,floatfix,aps]{revtex4-2}

\usepackage[latin9]{inputenc}
\usepackage{amsthm}
\usepackage{amsmath}
\usepackage{amssymb}
\usepackage{esint}
\usepackage{graphicx}
\usepackage{upgreek}

\makeatletter

\@ifundefined{textcolor}{}
{%
 \definecolor{BLACK}{gray}{0}
 \definecolor{WHITE}{gray}{1}
 \definecolor{RED}{rgb}{1,0,0}
 \definecolor{GREEN}{rgb}{0,1,0}
 \definecolor{BLUE}{rgb}{0,0,1}
 \definecolor{CYAN}{cmyk}{1,0,0,0}
 \definecolor{MAGENTA}{cmyk}{0,1,0,0}
 \definecolor{YELLOW}{cmyk}{0,0,1,0}
}

\usepackage{soul}
\usepackage{xspace}
\usepackage{braket}
\usepackage[backref=true,
bookmarksnumbered=true,
bookmarks=true,
bookmarksopen=true,
colorlinks=true,
citecolor=blue,
linkcolor=blue,
anchorcolor=green,
urlcolor=blue,unicode=false]{hyperref}

\let\baraccent=\= 
\renewcommand{\=}[1]{\stackrel{#1}{=}} 

\newcommand{\mos}{$\text{MoS}_\text{2}$}

\newcommand{\didv}{d$I$/d$V$}

\makeatother

\usepackage{babel}

\begin{document}

\title{
Tuning a two-impurity Kondo system by a moir\'e superstructure 
}

\author{Sergey Trishin}
\affiliation{\mbox{Fachbereich Physik, Freie Universit\"at Berlin, 14195 Berlin, Germany}}

\author{Christian Lotze}
\affiliation{\mbox{Fachbereich Physik, Freie Universit\"at Berlin, 14195 Berlin, Germany}}

\author{Friedemann Lohss}
\affiliation{\mbox{Fachbereich Physik, Freie Universit\"at Berlin, 14195 Berlin, Germany}}

\author{Giada Franceschi}
\affiliation{\mbox{Fachbereich Physik, Freie Universit\"at Berlin, 14195 Berlin, Germany}}

\author{Leonid I.\ Glazman}
\affiliation{Department of Physics, Yale University, New Haven, Connecticut 06520, USA}

\author{Felix von Oppen}
\affiliation{\mbox{Dahlem Center for Complex Quantum Systems and Fachbereich Physik, Freie Universit\"at Berlin, 14195 Berlin, Germany}}

\author{Katharina J. Franke}
\affiliation{\mbox{Fachbereich Physik, Freie Universit\"at Berlin, 14195 Berlin, Germany}}


\begin{abstract}
Two-impurity Kondo models are paradigmatic for correlated spin-fermion systems.  Working with Mn atoms on Au(111) covered by a monolayer of \mos, we tune the inter-adatom exchange via the adatom distance and the adatom-substrate exchange via the location relative to a moir\'e structure of the substrate. Differential-conductance measurements on isolated adatoms exhibit Kondo peaks with heights depending on the adatom location relative to the moir\'e structure. Mn dimers spaced by a few atomic lattice sites exhibit split Kondo resonances. In contrast, adatoms in closely spaced dimers couple antiferromagnetically, resulting in a molecular-singlet ground state. Exciting the singlet-triplet transition by tunneling electrons, we find that the singlet-triplet splitting is surprisingly sensitive to the moir\'e structure. We interpret our results theoretically by relating the variations in the singlet-triplet splitting to the heights of the Kondo peaks of single adatoms, finding evidence for coupling of the adatom spin to multiple conduction electron channels. 
\end{abstract}


\maketitle

Exchange interactions between magnetic adatoms and itinerant electrons of a substrate can induce correlation effects. For strong exchange coupling, the adatom spin becomes Kondo screened \cite{Madhavan1998, Li1998}. For intermediate coupling, where the Kondo temperature is comparable to temperature or other competing couplings, the Kondo renormalizations remain in the perturbative domain \cite{Zhang2013}. When the exchange coupling is weak, competing couplings such as single-ion anisotropy can dominate, in which case Kondo screening can be neglected and spin excitations can be probed \cite{Heinrich2004}. The panorama becomes yet broader when exchange coupling the adatom to a second magnetic atom in its vicinity. The nature of this coupling depends on the interatomic spacing. In close proximity, direct exchange tends to dominate, while larger separations favor substrate-mediated couplings such as the oscillatory Rudermann-Kittel-Kasuya-Yosida (RKKY) \cite{Ruderman1954, Kasuya1956, Yosida1957} and Dzyaloshinskii-Moriya (DM) \cite{Dzyaloshinsky1958, Moriya1960} interactions. The resulting ground states may be ferromagnetic \cite{Wahl2007,Meier2008}, antiferromagnetic \cite{Wahl2007,Meier2008}, or noncollinear \cite{Khajetoorians2016}. 

The competition between inter-adatom and adatom-substrate exchange leads to a rich phase diagram with multiple correlated ground states. Theoretically, the two-impurity Kondo problem has been treated extensively \cite{Jayaprakash1981,Jones1988}, and motivated numerous experiments
 \cite{Wahl2007,Tsukuhara2011,Prueser2014,Spinelli2015, Moro2019}.
The parameter space can be most directly explored by scanning-tunneling-microscope (STM) experiments. Atom manipulation with the STM tip admits manoevering the atoms into lattice sites at various distances and thus investigating different interatomic interaction strengths \cite{Khajetoorians2019}. Tuning of the exchange coupling to the surface is somewhat less straightforward. An early approach used strain-induced changes in the band gap of a decoupling interlayer \cite{Oberg2014}. A more controlled strategy would exploit well-defined superstructures. Prime candidates to impose a spatially periodic modulation of the atom--substrate interaction strength are interlayers which form moir\'e structures with the underlying metal substrate \cite{Ren2014, Jacobson2015, Trishin2021}. Most notably, monolayers of \mos\ on Au(111) have been successfully employed for tuning the exchange coupling of single magnetic Fe atoms from essentially uncoupled to strongly Kondo screened \cite{Trishin2021}. 

Here, we exploit the moir\'e pattern formed by monolayer-\mos\ on Au(111) to tune the exchange coupling of Mn dimers with the substrate, thereby probing the competition between interatomic and atom-substrate exchange. We find that the direct exchange coupling between closely-spaced Mn atoms leads to a singlet ground state and study the remarkably strong variations of the singlet-triplet splitting across the moir\'e pattern. 
We introduce a new experimental signature of multi-channel Kondo coupling by exploiting a theoretical relation between the singlet-triplet excitation energy of the dimer and the Kondo renormalizations of individual adatoms and find evidence that the Mn adatoms are coupled to several conduction-electron channels.

\begin{figure}[tb]
\includegraphics[width=0.95\columnwidth]{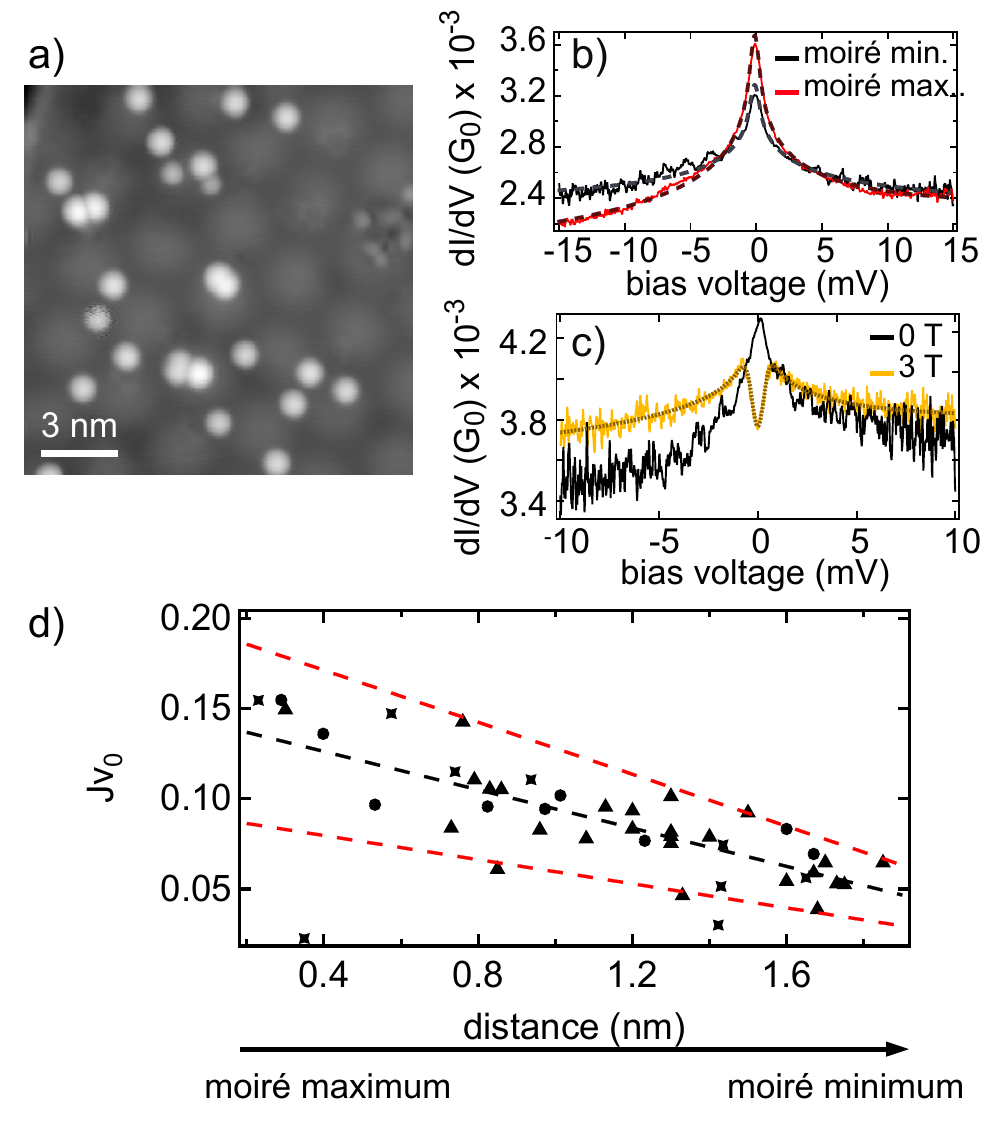}
\caption{Variation of the Kondo coupling across the moir\'e structure. a) STM topography of Mn atoms on the moir\'e structure of \mos\ on Au(111) (recorded at 100 mV, 20 pA). b) \didv spectra taken on Mn atoms adsorbed close to a moir\'e maximum (black) and a moir\'e minimum (red). The dashed lines show fits using a code based on Ref.\ \cite{Ternes2015}. The fits yield a (dimensionless) adatom-substrate exchange $J \nu_0$ of -0.080 (red) and -0.049 (black). c) \didv spectra on a Mn atom at 0 T (black) and 3 T (orange). The zero-bias resonance splits at 3 T (fit: dashed line). [Spectra were recorded at a setpoint of 15 mV, 3 nA (panel b) and 10 mV, 3 nA (panel c)]. d) Values of $J \nu_0$ obtained from fitting \didv spectra (keeping $T_{0}^{2}=0.000415$ constant, as obtained from a best fit with $B$-field) on atoms at various positions within the moir\'e superstructure (distance to the moir\'e maximum). Symbols indicate different measurement sets. The black dashed line is a linear guide to the eye through the data points. The red dashed lines are corresponding lines obtained from fits using different tunnel couplings $T_{0}^{2}$ . The upper line corresponds $T_{0}^{2}= 2.635 \times 10^{-4}$ and the lower line to $T_{0}^{2}=5.6 \times 10^{-4}$. These boundary values have been determined from error margins of fits at 3 T.}
\label{fig1}
\end{figure}

We use the previously established moir\'e-patterned decoupling layer \mos\ on Au(111) \cite{Trishin2021}, grown by depositing Mo atoms and subsequent annealing to 800 K in H$_2$S gas at a pressure $p=10^{-5}$\,mbar \cite{Gronborg2015, Krane2018a}. A moir\'e structure forms as a result of the lattice mismatch between adlayer and substrate, easily seen in the STM images as a modulation of the apparent height with a periodicity of $\approx 3.3$\,nm (Fig.\ \ref{fig1}a) \cite{Gronborg2015,Bana2018, Krane2018a}. Deposition of Mn atoms at low temperatures ($<10$\,K) leads to isolated atoms observable as round protrusions with an apparent height of $\approx 300$\,pm. Some round protrusions with smaller apparent height are attributed to Mn atoms attached to defects and excluded from further analysis. We also find some oval protrusions. As discussed in more detail below, we attribute these to Mn dimers. 

We start by characterizing individual Mn atoms. These exhibit a narrow zero-bias resonance in differential-conductance (\didv) spectra as shown for two examples in Fig.\ \ref{fig1}b. At our experimental temperature of 1.1\,K, the lineshape is well reproduced by a temperature-broadened logarithmic peak. The peak splits when applying a magnetic field (Fig.\ \ref{fig1}c). At 3\,T, the Zeeman split amounts to 600\,$\mu$V. This behavior is reminiscent of a weakly coupled Kondo impurity, with the experimental temperature larger than or of the order of the Kondo temperature \cite{Zhang2013}. Mn atoms at different positions with respect to the moir\'e lattice exhibit lineshapes with small variations in intensity, but the same broadening (see Fig.\ \ref{fig1}b for two extremal cases). The intensity modulations can be understood as modulations of $J\nu_0$, where $J$ is the strength of the exchange coupling to the conduction electrons and $\nu_0$ the density of states (DoS) at the Fermi level as discussed in more detail below. The observed variations are consistent with the DoS modulations due to the adatoms' position on the moir\'e structure (Fig.\ \ref{fig1}d). 
 
\begin{figure}[tb]
\includegraphics[width=0.95\columnwidth]{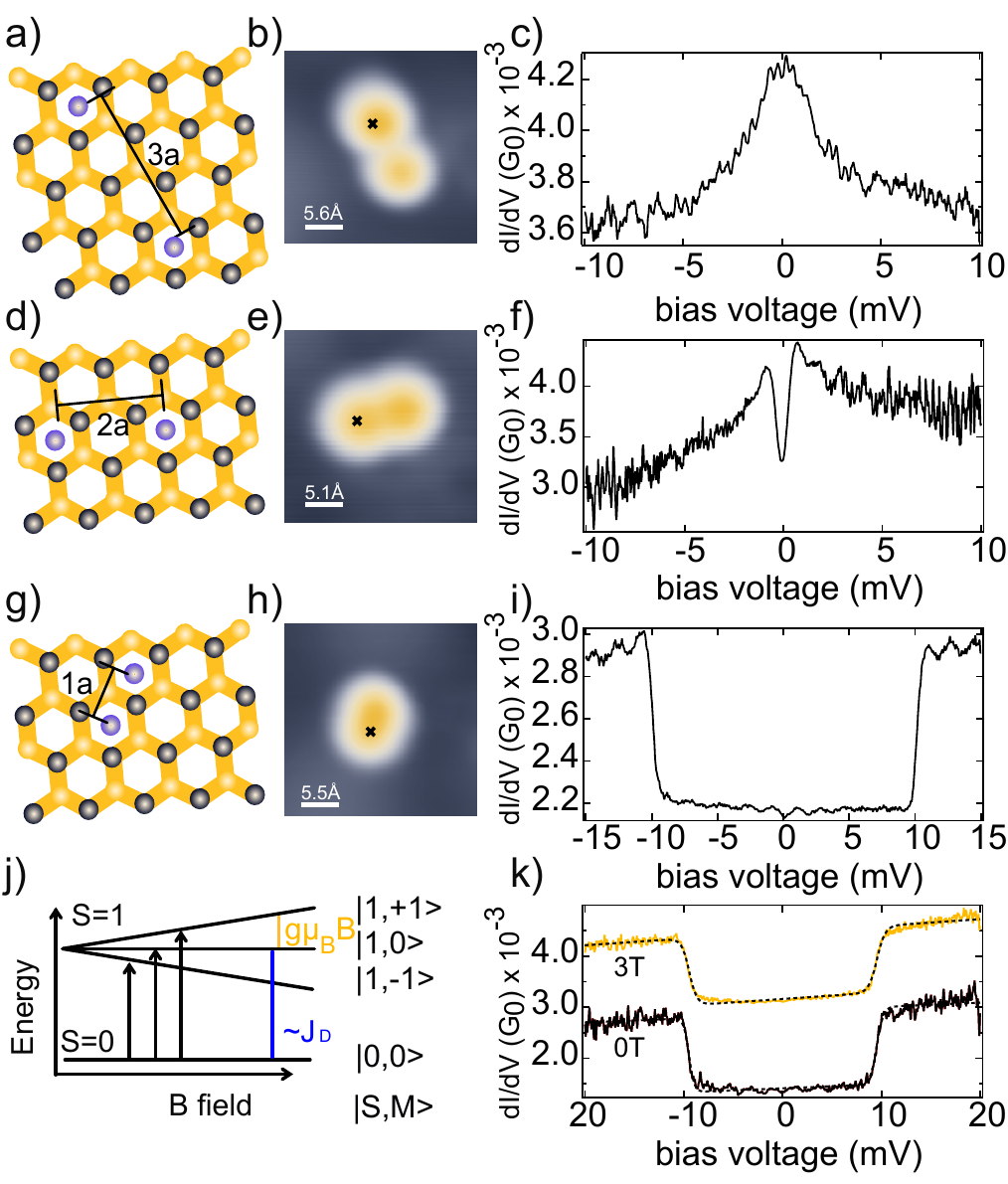}
\caption{Various dimer structures. a), d), g) Structure models and b), e), f) corresponding STM topographies of Mn dimers on \mos with various interatom spacings. Yellow, gray, and purple spheres represent S, Mo, and Mn, respectively. The Mn atoms, sitting in \mos\ hollow sites, are separated by three lattice sites (panels a,b), two lattice sites (panels d,e), and one lattice site (panels g,h). c), f), i) \didv spectra recorded at the locations indicated by the black crosses in (b,e,h). The spectra drastically depend on the dimer separation, exhibiting a Kondo resonance (panel c), a split Kondo resonance (panel f), and a step-like increase in the differential conductance (panel i). j) Energy-level diagram of the observed spin excitation in (i). The degeneracy of the $M=0, \pm 1$ sublevels of the excited state is lifted by a magnetic field. k) \didv spectra of a dimer with Mn in nearest-neighbor sites with and without magnetic field and respective fits with symmetric step functions (dashed). For our measurement conditions at 1.1 K, a magnetic field of 3 T is not sufficient to fully resolve the splitting, but the excitation appears broadened by 110 $\mu$V. STM topographies were recorded at 100 mV, 20 pA, the setpoint of the \didv spectra was 10 mV, 3 nA (c,f), 15 mV, 3 nA (i), and 20 mV, 3 nA (k).}
\label{fig2}
\end{figure}

Next, we characterize dimer structures formed by two adatoms in close proximity to each other. Density-functional calculations suggest that isolated atoms sit in hollow sites of the terminating S layer \cite{Wang2014}. Starting with this assumption, we can tentatively assign model structures to the most commonly found dimer arrangements on the surface by evaluating the separation and orientation in the STM images. Figure \ref{fig2}a,b shows an arrangement, where two Mn atoms are separated by three lattice sites of the \mos\ substrate. At this separation, the atoms show a Kondo resonance as previously described for individual adatoms (Fig.\ \ref{fig2}c), indicating that interatomic interactions are negligible. At a distance of two atomic lattice sites (Fig.\ \ref{fig2}d,e), the Kondo resonance develops a dip at the Fermi level (Fig.\ \ref{fig2}f). The spectrum is reminiscent of a Zeeman-split Kondo resonance, indicating magnetic interactions between the atoms, presumably resulting from substrate-mediated RKKY interactions. When the atoms are in even closer proximity, their shapes are no longer individually resolved in the STM image (Fig.\ \ref{fig2}g,h). While a definite assignment of the adsorption sites is thus difficult, the oval shape and its orientation with respect to the underlying lattice suggest that the atoms lie in nearest-neighbor hollow sites (for details and STM manipulations, see section S2 in Supplementary Material (SM) \cite{SM}). Differential-conductance spectra measured on this type of dimer are radically different from those of individual atoms or weakly interacting dimers. The Kondo resonance is now replaced by pronounced inelastic steps at $\pm$10\,mV (Fig.\ \ref{fig2}i). 

It is rather surprising to detect inelastic excitations of a relatively large energy, considering that individual atoms do not show a noticeable magnetocrystalline anisotropy. Instead of a change in magnetocrystalline anisotropy energy, we suggest that the threshold energy is associated with a spin-changing transition of the dimer. Such excitations have been observed for Mn dimers on CuN \cite{Hirjibehedin2006}. The close proximity of the atoms may allow for direct exchange as a result of finite overlap of the atomic $d$ orbitals. Mn atoms are indeed likely coupled antiferromagnetically when interacting via direct exchange \cite{Mokrousov2007}. This would lead to a singlet ground state $\left|S_\mathrm{tot}=0, M=0\right\rangle$. Magnetic excitations must then involve a spin-changing transition such as the singlet-triplet transition and the excitation energy directly reflects the exchange coupling $J_D$ (for details, see below). To further corroborate the antiferromagnetic nature of the exchange coupling, we apply an external magnetic field of 3\,T to the dimer. The inelastic steps become slightly broader (Fig.\ \ref{fig2}k). This is consistent with a singlet-triplet transition to $\left|S_\mathrm{tot}=1, M \right\rangle$, where the excited state is Zeeman split in the magnetic field, but the sublevels are not individually resolved at the experimental temperature of 1.1\,K (Fig.\ \ref{fig2}j). Importantly, we do not observe additional excitations around zero bias, which would indicate a higher-spin ground state as favored by ferromagnetic coupling of the atoms. 

\begin{figure}[tb]
\includegraphics[width=0.95\columnwidth]{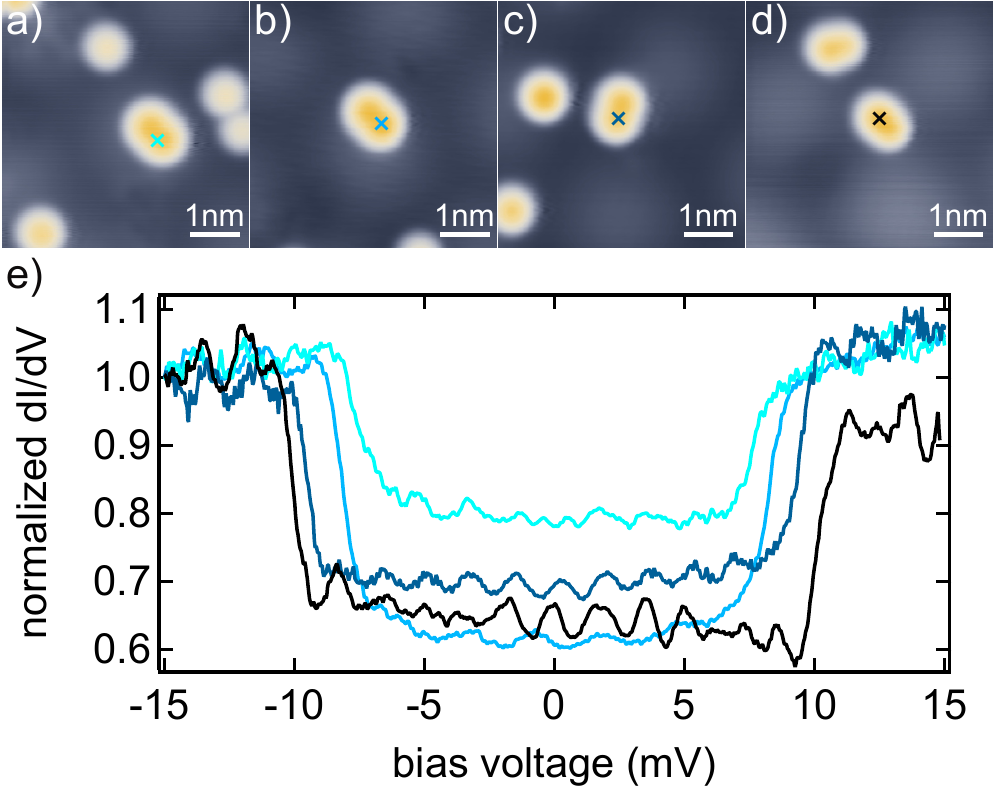}
\caption{Antiferromagnetically coupled Mn dimers (oval structures), which are identical apart from their location relative to the moir\'e structure. a-d) STM topographies of dimers (a) on the maximum, (b) close to the maximum, (c) further from the maximum, and (d) at the minimum of the moir\'e structure. e) \didv spectra acquired on the dimers shown in (a-d), with colors matched to the crosses in (a-d). Topographies recorded at 100 mV, 20 pA, setpoints of the recorded spectra were 20 mV, 1 nA (b), 20 mV, 3 nA (a) and 15 mV, 3 nA (c), (d). Spectra are normalized for clarity.
}
\label{fig3}
\end{figure}

As discussed above, the moir\'e superstructure weakly affects the height of the Kondo resonance of individual atoms reflecting the modulation of the dimensionless exchange coupling $J\nu_0$. As the dimer is in a singlet ground state, one may naively expect that the moir\'e structure does not influence the inelastic excitations. Remarkably, we observe strong variations of the singlet--triplet transition by several meV as the dimer's adsorption site is varied with respect to the moir\'e lattice (Fig.\ \ref{fig3}). Dimers located on maxima of the moir\'e structure (Fig.\ \ref{fig3}a,e) exhibit the smallest excitation energy (7.5\,meV), while those on minima (Fig.\ \ref{fig3}d,e) show the largest excitation energy (10\,meV). 

To understand these variations, we compute the shift of the singlet-triplet splitting $\Delta$ due to the hybridization of the adatom $d$ orbitals with the substrate and relate it to the exchange coupling between the adatom and conduction-electron spins. As we do not observe inelastic excitations on single adatoms indicating negligible single-ion anisotropy, we assume that the Mn atoms are only weakly perturbed by the surrounding and retain the half-filled $d$-shell when placed on the substrate. According to Hund's rule, this implies a high-spin configuration with $S=5/2$ and suggests that spin-orbit coupling will be weak, so that the inter-adatom exchange can be modeled by isotropic Heisenberg exchange, $H_\mathrm{ex}=J_D\mathbf{S}_{A}\cdot\mathbf{S}_{B}$. Here, $\mathbf{S}_{A,B}$ denotes the spins of adatom A,B. 

In the absence of hybridization with the substrate, states with magnitude $S_\mathrm{tot}$ of the total spin $\mathbf{S}_\mathrm{tot}=\mathbf{S}_A+\mathbf{S}_B$ will have direct exchange energy
\begin{equation}
   E_\mathrm{ex}(S_A,S_B; S_\mathrm{tot}) = \frac{J_D}{2}[S_\mathrm{tot}
                          (S_\mathrm{tot}+1)-\!\!\!\!\sum_{j\in\{A,B\}} \!\! S_j(S_j+1)].
\end{equation}
Evaluating the singlet-triplet splitting for $S_A,S_B=\frac{5}{2}$, we find 
\begin{equation}
   \Delta = E_\mathrm{ex}(\frac{5}{2},\frac{5}{2}; 1) - E_\mathrm{ex}(\frac{5}{2},\frac{5}{2};0) = J_D.
\end{equation}
This splitting is reduced by the hybridization of the adatom $d$ orbitals with the conduction electrons. In general, the $d$ orbitals hybridize with $2S=5$ (symmetry-adapted) conduction-electron channels \cite{Schrieffer1967}. Since the substrate breaks rotational symmetry, the strength of hybridization $V_m$ depends on the channel $m$. The energies of the singlet and triplet states are then shifted by virtual excitation processes, in which a $d$ electron hops into the substrate or a substrate electron hops into the $d$ shell. Physically, these processes reduce the effective adatom spin, which results in a smaller direct exchange. A detailed calculation in second-order perturbation theory (see section S1 in SM for details \cite{SM}) gives a renormalized singlet-triplet splitting
\begin{equation}
    \Delta  = J_D\left\{ 1 - \frac{2}{5} \sum_m  \nu_0 |V_m|^2 \left[ \frac{1}{  |
    \epsilon_d| }  +   \frac{1}{\epsilon_d + U } \right] \right\}.
\end{equation}
Here, $-\epsilon_d>0$ is the energy to remove an electron from the filled $d$-shell and $\epsilon_d +U$ the energy to add an electron. The factor $2$ in front of the sum over channels accounts for the fact that both adatoms can be excited. The factor $1/5$ results from angular-momentum coupling. 

The singlet-triplet spacing can be directly related to experimentally measurable quantities by noting that the exchange coupling between the conduction electrons and the spin-$S$ adatom is given by \cite{Schrieffer1967}
\begin{equation}
    J_m  =   \frac{\nu_0 |V_m|^2}{2S}  \left[ \frac{1}{  |
    \epsilon_d| }  +   \frac{1}{\epsilon_d + U } \right] ,
\end{equation}
so that we can express the singlet-triplet splitting of the $S=\frac{5}{2}$ Mn dimer as 
\begin{equation}
    \Delta  = J_D\left\{ 1 - 2 \sum_m \nu_0 J_m  \right\}.
    \label{eq:singtrip}
\end{equation}
For weak coupling ($\nu_0 J_m\ll 1$), the relative change in the singlet-triplet spacing between minimum ($\Delta_\mathrm{min}$) and maximum ($\Delta_\mathrm{max}$) is approximately equal to $\delta \simeq (\Delta_\mathrm{min}-\Delta_\mathrm{max})/J_D$. Equation (\ref{eq:singtrip}) relates this directly to the corresponding change in the sum of the dimensionless exchange couplings $\sum_m \nu_0 J_m$ to the substrate. Since information on the exchange couplings $\nu_0 J_m$ can be extracted from the Kondo data on a single adatom, applying this relation to the data in Fig.\ \ref{fig3}e gives direct information on the number of conduction-electron channels coupled to the adatom spins. 

For analyzing the number of participating channels, we first assume that the adatom spin is coupled to a single channel. With this assumption, we can extract the dimensionless adatom-substrate exchange coupling $\nu_0 J$ of the single channel by fitting the Kondo peak of the isolated atoms using a program based on Ref.\ \cite{Ternes2015}. Showcasing the variation between extremal positions with respect to the moir\'e pattern, we extracted a value of $\nu_0 J=0.049$ for the adatom on the moir\'e minimum and $\nu_0 J=0.080$ for an atom on the maximum from fitting the Kondo data in Fig.\ \ref{fig1}b. Equation \ref{eq:singtrip} (specified to a single channel) then predicts a relative change $\delta$ of the singlet-triplet splitting from minimum to maximum by $\approx$ 6\%. This is clearly smaller than the experimentally observed variation of $\approx 25$\% (Fig.\ \ref{fig3}e). We have extracted $\nu_0 J$ for several dozen isolated atoms in various positions across the moir\'e structure. In all cases, $\nu_0 J$ decreases with increasing distance from the maxima of the moir\'e pattern (Fig.\ \ref{fig1}d). The variations of $\nu_0 J$ for similar distances from the moir\'e maxima partially derive from the lack of rotational symmetry, so that the distance to the moir\'e maximum does not uniquely specify the adsorption site. Moreover, the fitting procedure contains some uncertainty, as the strength of tunneling $T_0^2$ and $\nu_0 J$ both affect the peak height. We first determined $T_0^2$ from a spectrum of an atom subject to a magnetic field (for which the uncertainty is reduced due to the additional magnetic-field induced structure). We then fitted all spectra with the extracted value of $T_0^2$. To indicate the error margins of the fits, we reran all fits taking extremal values of $T_0^2$ consistent with the $B$-field data with sufficient accuracy. The black dashed line in Fig.\ \ref{fig1}d shows a linear guide-to-the-eye for the best fit results, while the red dashed lines indicate the scalings obtained when using the extremal values of $T_0^2$. We find that with the assumption of a single channel, only the largest variation in $\nu_0 J$ (upper red dashed line) would explain the variation of the singlet-triplet splitting.

We can also apply Eq.\ (\ref{eq:singtrip}), when assuming that all five channels are equally coupled. Since each channel renormalizes independently, the value of $\nu_0 J_m$ for any $m$ is equal to that extracted with the single-channel assumption. (In the leading-logarithm approximation underlying the Kondo fits, the number of channels enters only as an overall prefactor, which can be absorbed into $T_0^2$.) With this assumption, Eq.\ (\ref{eq:singtrip}) predicts a variation in the singlet-triplet spacing, which is larger by a factor of five than in the single-channel case. We then find that the observed variation in the singlet-triplet spacing across the moir\'e structure (Fig.\ \ref{fig3}e) would only be consistent with the opposite extreme case (lower red dashed line in Fig.\ \ref{fig1}d). Thus, while the uncertainties of the fitting procedure preclude a fully quantitative analysis, 
our results strongly suggest that the Mn atoms have substantial coupling to several conduction-electron channels in the Au(111) substrate. 

In conclusion, we varied the adatom-substrate exchange of Mn monomers and dimers by exploiting the moir\'e pattern of a \mos\ layer on Au(111). The moir\'e structure imprints density-of-states modulations, which in turn affect the Kondo resonance of the monomer and the singlet-triplet splitting of antiferromagnetically coupled dimers. Relating these variations through a theoretical analysis, we find evidence that the adatoms are coupled to multiple conduction-electron channels. This constrasts with the commonly made assumption that adatoms couple only to a single channel of a metallic substrate. Our results show that this assumption is violated in the perturbative limit. For the fully developed Kondo effect, relatively small differences in $\nu_0 J_m$ between channels result in large differences in the associated Kondo temperatures $T_{K,m} \propto e^{-{1}/{\nu_0 J_m}}$. Then, the single-channel approximation can still be adequate provided that only the first stage of the resulting multistage Kondo screening is accessible in experiment. Interestingly, coupling to multiple conduction-electron channels has previously been invoked to explain the appearance of multiple Yu-Shiba-Rusinov states induced by magnetic adatoms on superconductors \cite{Ruby2016,Choi2017}. Our results emphasize that adatom dimers realize a rich two-impurity problem. While theoretical studies have focused on spin-$\frac{1}{2}$ impurities, adatom dimers typically have higher spins and couple to multiple conduction-electron channels. 

\begin{acknowledgments}
We acknowledge financial support by the Deutsche Forschungsgemeinschaft (DFG, German Research Foundation) through project numbers 328545488 (CRC 227, project B05) and  277101999 (CRC 183, project C02 and a Mercator professorship), as well as by the National Science Foundation through grant  NSF DMR-2002275.
\end{acknowledgments}

%

\clearpage

\renewcommand{\figurename}{Supplementary Figure}
\renewcommand{\figurename}{Supplementary Figure}
\renewcommand{\tablename}{Supplementary Table}

\setcounter{figure}{0}
\setcounter{section}{0}
\setcounter{equation}{0}
\setcounter{table}{0}

\onecolumngrid

\newcommand{\vsigma}{\mbox{\boldmath $\sigma$}}

\section*{\Large{Supplementary Material}}

\maketitle 

\section{Theoretical considerations}

We provide details concerning the theoretical considerations in the main text. We assume that Mn retains its half-filled $d$ shell in the presence of the weak coupling to the substrate. The uncoupled state of Mn is thus fully rotationally symmetric and coupled to five conduction-electron channels. As the rotational symmetry is broken by the coupling to the substrate, their hybridization $V_m$ with the various conduction-electron channels will be different. In the following, we compute the singlet-triplet splitting perturbatively, focusing on one channel ($m=0$ for definiteness). The general result is obtained by adding the independent corrections for all five channels. 

\subsection{Spin states of monomer}

First consider the spin states of a single Mn adatom. We can generate the spin states
$|\frac{5}{2},S_z\rangle$ by applying the spin lowering operator  $S_- = \sum_{m=-2}^2 c_{m,\downarrow}^\dagger c_{m,\uparrow}^{\phantom{\dagger}}$ to
\begin{equation}
  |\frac{5}{2},\frac{5}{2}\rangle = \prod_m c^\dagger_{m,\uparrow} |\mathrm{vac}
  \rangle .
\end{equation}
Then, we have 
\begin{eqnarray}
|\frac{5}{2},\frac{5}{2}\rangle &=& |\uparrow \uparrow \uparrow \uparrow \uparrow \rangle
\nonumber\\
   |\frac{5}{2},\frac{3}{2}\rangle &=& \sqrt\frac{1}{5} \sum |\textrm{states with one 
   flipped spin}\rangle \nonumber\\    
   |\frac{5}{2},\frac{1}{2}\rangle &=& \sqrt\frac{1}{10}\sum |\textrm{states with two 
   flipped spins}\rangle \nonumber\\    
   |\frac{5}{2},-\frac{1}{2}\rangle &=& \sqrt\frac{1}{10}\sum |\textrm{states with three 
   flipped spins}\rangle \nonumber\\    
      |\frac{5}{2},-\frac{3}{2}\rangle &=& \sqrt\frac{1}{5}\sum |\textrm{states with four 
   flipped spins}\rangle \nonumber\\
      |\frac{5}{2},-\frac{5}{2}\rangle &=& |\downarrow \downarrow 
      \downarrow \downarrow \downarrow \rangle .
\end{eqnarray}
Similarly, we can derive the states with one less electron, say in the $m=0$ state. One finds
\begin{eqnarray}
|2,2\rangle &=& |\uparrow \uparrow \uparrow \uparrow \rangle
\nonumber\\
   |2,1\rangle &=& \sqrt\frac{1}{4}\sum |\textrm{states with one 
   flipped spin}\rangle \nonumber\\    
   |2,0\rangle &=& \sqrt\frac{1}{6}\sum |\textrm{states with two 
   flipped spins}\rangle \nonumber\\    
   |2,-1\rangle &=& \sqrt\frac{1}{4}\sum |\textrm{states with three 
   flipped spins}\rangle \nonumber\\    
      |2,-2\rangle &=& |\downarrow 
      \downarrow \downarrow \downarrow \rangle .
\end{eqnarray}
Applying $c_{0,\uparrow}$ to the $S=\frac{5}{2}$ states, one finds
\begin{eqnarray}
c_{0,\uparrow}|\frac{5}{2},\frac{5}{2}\rangle &=& |2,2 \rangle
\nonumber\\
   c_{0,\uparrow}|\frac{5}{2},\frac{3}{2}\rangle &=& \sqrt\frac{4}{5}  |2, 1 \rangle \nonumber\\    
   c_{0,\uparrow}|\frac{5}{2},\frac{1}{2}\rangle &=& \sqrt\frac{6}{10} |2,0 \rangle \nonumber\\    
   c_{0,\uparrow}|\frac{5}{2},-\frac{1}{2}\rangle &=& \sqrt\frac{4}{10} |2,-1\rangle \nonumber\\    
     c_{0,\uparrow} |\frac{5}{2},-\frac{3}{2}\rangle &=& \sqrt\frac{1}{5} |2,-2\rangle \nonumber\\
     c_{0,\uparrow} |\frac{5}{2},-\frac{5}{2}\rangle &=& 0.
\end{eqnarray}
Applying $c_{0,\downarrow}$ to the $S=\frac{5}{2}$ states, one finds
\begin{eqnarray}
c_{0,\downarrow}|\frac{5}{2},\frac{5}{2}\rangle &=& 0
\nonumber\\
   c_{0,\downarrow}|\frac{5}{2},\frac{3}{2}\rangle &=& \sqrt\frac{1}{5}  |2, 2 \rangle \nonumber\\    
   c_{0,\downarrow}|\frac{5}{2},\frac{1}{2}\rangle &=& \sqrt\frac{4}{10} |2,1 \rangle \nonumber\\    
   c_{0,\downarrow}|\frac{5}{2},-\frac{1}{2}\rangle &=& \sqrt\frac{6}{10} |2,0\rangle \nonumber\\    
     c_{0,\downarrow} |\frac{5}{2},-\frac{3}{2}\rangle &=& \sqrt\frac{4}{5} |2,-1\rangle \nonumber\\
     c_{0,\downarrow} |\frac{5}{2},-\frac{5}{2}\rangle &=& |2,-2\rangle.
\end{eqnarray}

\subsection{Singlet state of dimer -- tunneling out}

The spin state of the dimer can either be expanded in product states $|S_1,M_1\rangle\otimes |S_2,M_2\rangle$ of the two adatoms, or according to magnitude $S_\mathrm{tot}$ and projection $M_\mathrm{tot}$ of the total angular momentum $\mathbf{S}_\mathrm{tot}=\mathbf{S}_1+\mathbf{S}_2$ as $|S_1,S_2;S_\mathrm{tot},M_\mathrm{tot}\rangle$. First consider the singlet state of the dimer. Using Clebsch-Gordan coefficients, we can expand it into product states as
\begin{eqnarray}
   |\frac{5}{2},\frac{5}{2}; 0,0\rangle &=& \sqrt\frac{1}{6} \left( 
   |\frac{5}{2},\frac{5}{2} \rangle \otimes | \frac{5}{2},-\frac{5}{2} \rangle
   -    |\frac{5}{2},\frac{3}{2}\rangle \otimes |\frac{5}{2},-\frac{3}{2} \rangle
   +    |\frac{5}{2},\frac{1}{2}\rangle \otimes |\frac{5}{2},-\frac{1}{2} \rangle
    \right.\nonumber\\
  &&   \qquad \left. -     |\frac{5}{2},-\frac{1}{2}\rangle \otimes |\frac{5}{2},\frac{1}{2} \rangle
   +    |\frac{5}{2},-\frac{3}{2}\rangle \otimes |\frac{5}{2},\frac{3}{2} \rangle
    -    |\frac{5}{2},-\frac{5}{2}\rangle \otimes |\frac{5}{2},\frac{5}{2}\rangle 
   \right).
\end{eqnarray}
Applying $c_{L,0,\uparrow}$ for the left adatom gives
\begin{eqnarray}
   c_{L,0,\uparrow}|\frac{5}{2},\frac{5}{2}; 0,0\rangle &=& \sqrt\frac{1}{6} \left( 
        |2,2 \rangle \otimes | \frac{5}{2},-\frac{5}{2} \rangle
   -    \sqrt\frac{4}{5}|2,1\rangle \otimes |\frac{5}{2},-\frac{3}{2} \rangle
   +    \sqrt\frac{6}{10}|2,0\rangle \otimes |\frac{5}{2},-\frac{1}{2} \rangle
    \right.\nonumber\\
  &&   \qquad \left. -   \sqrt\frac{4}{10}  |2,-1\rangle \otimes |\frac{5}{2},\frac{1}{2} 
  \rangle
   +   \sqrt\frac{1}{5} |2,-2\rangle \otimes |\frac{5}{2},\frac{3}{2} \rangle
   \right)
\end{eqnarray}
Similarly, we have
\begin{eqnarray}
   c_{L,0,\downarrow}|\frac{5}{2},\frac{5}{2}; 0,0\rangle &=& \sqrt\frac{1}{6} \left( 
       - \sqrt\frac{1}{5}|2,2 \rangle \otimes | \frac{5}{2},-\frac{3}{2} \rangle
   +    \sqrt\frac{4}{10}|2,1\rangle \otimes |\frac{5}{2},-\frac{1}{2} \rangle
   -    \sqrt\frac{6}{10}|2,0\rangle \otimes |\frac{5}{2},\frac{1}{2} \rangle
    \right.\nonumber\\
  &&   \qquad \left. +   \sqrt\frac{4}{5}  |2,-1\rangle \otimes |\frac{5}{2},\frac{3}{2} 
  \rangle
   -    |2,-2\rangle \otimes |\frac{5}{2},\frac{5}{2} \rangle
   \right)
\end{eqnarray}
We can compare these states to 
\begin{eqnarray}
   |2,\frac{5}{2}; \frac{1}{2},\frac{1}{2}\rangle &=&  
   \sqrt\frac{1}{15} |2,2 \rangle \otimes | \frac{5}{2},-\frac{3}{2} \rangle
   -   \sqrt\frac{2}{15} |2,1\rangle \otimes |\frac{5}{2},-\frac{1}{2} \rangle
   +    \sqrt\frac{1}{5} |2,0\rangle \otimes |\frac{5}{2},\frac{1}{2} \rangle
    \nonumber\\
  &&   \qquad  
    -    \sqrt\frac{4}{15} |2,-1\rangle \otimes |\frac{5}{2},\frac{3}{2} \rangle
   +    \sqrt\frac{1}{3} |2,-2\rangle \otimes |\frac{5}{2},\frac{5}{2} \rangle
   \\
   |2,\frac{5}{2}; \frac{1}{2},-\frac{1}{2}\rangle &=&  
   \sqrt\frac{1}{3} |2,2 \rangle \otimes | \frac{5}{2},-\frac{5}{2} \rangle
   -   \sqrt\frac{4}{15} |2,1\rangle \otimes |\frac{5}{2},-\frac{3}{2} \rangle
   +    \sqrt\frac{1}{5} |2,0\rangle \otimes |\frac{5}{2},-\frac{1}{2} \rangle
    \nonumber\\
  &&   \qquad  
    -    \sqrt\frac{2}{15} |2,-1\rangle \otimes |\frac{5}{2},\frac{1}{2} \rangle
   +    \sqrt\frac{1}{15} |2,-2\rangle \otimes |\frac{5}{2},\frac{3}{2} \rangle ,
\end{eqnarray}
so that we identify
\begin{equation}
    c_{L,0,\uparrow}|\frac{5}{2},\frac{5}{2}; 0,0\rangle = -\sqrt\frac{1}{2}|2,\frac{5}{2}; \frac{1}{2},-\frac{1}{2}\rangle 
      \qquad ; \qquad 
       c_{L,0,\downarrow}|\frac{5}{2},\frac{5}{2}; 0,0\rangle
         = \sqrt\frac{1}{2} |2,\frac{5}{2}; \frac{1}{2},\frac{1}{2}\rangle .
\end{equation}

\subsection{Singlet state of dimer -- tunneling in}

Now consider tunneling in of an electron. We can follow the same steps. Now, the $m=0$ state of one of the atoms will be doubly occupied rather than empty, but this is also a zero-spin state. Thus, all the Clebsch-Gordan coefficients remain the same and one finds
\begin{equation}
    c^\dagger_{L,0,\uparrow}|\frac{5}{2},\frac{5}{2}; 0,0\rangle = \sqrt\frac{1}{2}|2,\frac{5}{2}; \frac{1}{2},\frac{1}{2}\rangle 
      \qquad ; \qquad 
       c^\dagger_{L,0,\downarrow}|\frac{5}{2},\frac{5}{2}; 0,0\rangle
         = -\sqrt\frac{1}{2} |2,\frac{5}{2}; \frac{1}{2},-\frac{1}{2}\rangle .
\end{equation}

\subsection{Triplet state of dimer -- tunneling out}

We expand the triplet state of the dimer into product states of the two monomers. Due to rotational invariance, we can consider the $M=1$ state without loss of generality,  
\begin{eqnarray}
   |\frac{5}{2},\frac{5}{2}; 1,1\rangle &=& 
  \sqrt\frac{1}{7}  |\frac{5}{2},\frac{5}{2} \rangle \otimes | \frac{5}{2},-\frac{3}{2} \rangle
   -  \sqrt\frac{8}{35}  |\frac{5}{2},\frac{3}{2}\rangle \otimes |\frac{5}{2},-\frac{1}{2} \rangle
   + \sqrt\frac{9}{35}   |\frac{5}{2},\frac{1}{2}\rangle \otimes |\frac{5}{2},\frac{1}{2} \rangle
    \nonumber\\
  &&   \qquad  -  \sqrt\frac{8}{35}   |\frac{5}{2},-\frac{1}{2}\rangle \otimes |\frac{5}{2},\frac{3}{2} \rangle
   +  \sqrt\frac{1}{7}  |\frac{5}{2},-\frac{3}{2}\rangle \otimes |\frac{5}{2},\frac{5}{2} \rangle.
\end{eqnarray}
Applying $c_{L,0,\uparrow}$ for the left adatom gives
\begin{eqnarray}
   c_{L,0,\uparrow}|\frac{5}{2},\frac{5}{2}; 1,1\rangle &=&   \sqrt\frac{1}{7}  |2,2 \rangle \otimes | \frac{5}{2},-\frac{3}{2} \rangle
   -  \sqrt\frac{8}{35}  \sqrt\frac{4}{5} |2,1\rangle \otimes |\frac{5}{2},-\frac{1}{2} \rangle
   + \sqrt\frac{9}{35} \sqrt\frac{6}{10}  |2,0\rangle \otimes |\frac{5}{2},\frac{1}{2} \rangle
    \nonumber\\
  &&   \qquad  - \sqrt\frac{8}{35} \sqrt\frac{4}{10}   |2,-1\rangle \otimes |\frac{5}{2},\frac{3}{2} \rangle
   +  \sqrt\frac{1}{7} \sqrt\frac{1}{5} |2,-2\rangle \otimes |\frac{5}{2},\frac{5}{2} \rangle.
\end{eqnarray}
Similarly,
\begin{eqnarray}
   c_{L,0,\downarrow}|\frac{5}{2},\frac{5}{2}; 1,1\rangle &=&  - \sqrt\frac{8}{35} 
   \sqrt\frac{1}{5} |2,2 \rangle \otimes | \frac{5}{2},-\frac{1}{2} \rangle
   +  \sqrt\frac{9}{35}  \sqrt\frac{4}{10} |2,1\rangle \otimes |\frac{5}{2},\frac{1}{2} \rangle
   - \sqrt\frac{8}{35} \sqrt\frac{6}{10}  |2,0\rangle \otimes |\frac{5}{2},\frac{3}{2} \rangle
    \nonumber\\
  &&   \qquad  + \sqrt\frac{1}{7} \sqrt\frac{4}{5}   |2,-1\rangle \otimes |\frac{5}{2},\frac{5}{2} \rangle.
\end{eqnarray}
We can compare this to
\begin{eqnarray}
   |2,\frac{5}{2}; \frac{1}{2},\frac{1}{2}\rangle &=&   \sqrt\frac{1}{15} 
    |2,2 \rangle \otimes | \frac{5}{2},-\frac{3}{2} \rangle
   -  \sqrt\frac{2}{15}   |2,1\rangle \otimes |\frac{5}{2},-\frac{1}{2} \rangle
   + \sqrt\frac{1}{5}   |2,0\rangle \otimes |\frac{5}{2},\frac{1}{2} \rangle
   \nonumber\\
   &&\qquad\qquad\qquad
     - \sqrt\frac{4}{15}   |2,-1\rangle \otimes |\frac{5}{2},\frac{3}{2} \rangle
     + \sqrt\frac{1}{3}   |2,-2\rangle \otimes |\frac{5}{2},\frac{5}{2} \rangle
   \nonumber\\
      |2,\frac{5}{2}; \frac{3}{2},\frac{1}{2}\rangle &=&   \sqrt\frac{32}{105} 
    |2,2 \rangle \otimes | \frac{5}{2},-\frac{3}{2} \rangle
   -  \sqrt\frac{5}{21}   |2,1\rangle \otimes |\frac{5}{2},-\frac{1}{2} \rangle
   + \sqrt\frac{2}{35}   |2,0\rangle \otimes |\frac{5}{2},\frac{1}{2} \rangle
   \nonumber\\
   &&\qquad\qquad\qquad
     + \sqrt\frac{2}{105}   |2,-1\rangle \otimes |\frac{5}{2},\frac{3}{2} \rangle
     - \sqrt\frac{8}{21}   |2,-2\rangle \otimes |\frac{5}{2},\frac{5}{2} \rangle
   \nonumber\\
   |2,\frac{5}{2}; \frac{3}{2},\frac{3}{2}\rangle &=&   \sqrt\frac{4}{35} 
    |2,2 \rangle \otimes | \frac{5}{2},-\frac{1}{2} \rangle
   -  \sqrt\frac{9}{35}   |2,1\rangle \otimes |\frac{5}{2},\frac{1}{2} \rangle
   + \sqrt\frac{12}{35}   |2,0\rangle \otimes |\frac{5}{2},\frac{3}{2} \rangle
     - \sqrt\frac{2}{7}   |2,-1\rangle \otimes |\frac{5}{2},\frac{5}{2} \rangle,
\end{eqnarray}
so that we identify
\begin{equation}
   c_{L,0,\uparrow}|\frac{5}{2},\frac{5}{2}; 1,1\rangle = \sqrt\frac{7}{15} |2,\frac{5}{2}; \frac{1}{2},\frac{1}{2}\rangle
     + \sqrt\frac{2}{15} |2,\frac{5}{2}; \frac{3}{2},\frac{1}{2}\rangle
   \qquad ; \qquad
    c_{L,0,\downarrow}|\frac{5}{2},\frac{5}{2}; 1,1\rangle =  - \sqrt\frac{2}{5} |2,\frac{5}{2}; \frac{3}{2},\frac{3}{2}\rangle 
\end{equation}

\subsection{Triplet state of dimer -- tunneling in}

This follows again by analogy with the tunneling-out terms, so that
\begin{equation}
   c^\dagger_{L,0,\downarrow}|\frac{5}{2},\frac{5}{2}; 1,1\rangle = \sqrt\frac{7}{15} |2,\frac{5}{2}; \frac{1}{2},\frac{1}{2}\rangle
     + \sqrt\frac{2}{15} |2,\frac{5}{2}; \frac{3}{2},\frac{1}{2}\rangle
   \qquad ; \qquad
    c^\dagger_{L,0,\uparrow}|\frac{5}{2},\frac{5}{2}; 1,1\rangle =  - \sqrt\frac{2}{5} |2,\frac{5}{2}; \frac{3}{2},\frac{3}{2}\rangle 
\end{equation}

\subsection{Singlet-triplet splitting}

In the absence of coupling to the substrate, the impurity spins $\mathbf{S}_1$ and $\mathbf{S}_2$ of the two Mn adatoms are subject to antiferromagnetic exchange coupling of the dimer, $H_\mathrm{ex} = J_D \mathbf{S}_1\cdot \mathbf{S}_2$ with $J_D>0$. Depending on the total spin $S_\mathrm{tot}$, the coupling energy is 
\begin{equation}
   E_\mathrm{ex}(S_1,S_2; S_\mathrm{tot}) = \frac{J_D}{2}  [S_\mathrm{tot}(S_\mathrm{tot}+1) - S_1(S_1+1)-S_2(S_2+1)].
\end{equation}
For Mn adatoms with $S_1=S_2=\frac{5}{2}$, the splitting between the triplet ($S=1$) excited state and the singlet ($S=0$) ground state is equal to $\Delta E^{(0)}_\mathrm{st} = J_D$. 

The singlet-triplet splitting is renormalized due the coupling of the adatoms to the substrate electrons. Tunneling of electrons between adatom $d$ orbitals and substrate couples the singlet to the intermediate states $|2,\frac{5}{2}; \frac{1}{2},\pm\frac{1}{2}\rangle$. The singlet state has exchange energy
\begin{equation}
   E_\mathrm{ex}(\frac{5}{2},\frac{5}{2}; 0) =  - \frac{35 J_D}{4}  ,
\end{equation}
while the intermediate states have exchange energy
\begin{equation}
   E_\mathrm{ex}(2,\frac{5}{2}; \frac{1}{2}) =  - 7 J_D  .
\end{equation}
In the absense of hybridization, we can then write the energy of of singlet state as 
\begin{equation}
   E_s^{(0)}  = 2 E_\mathrm{Mn} + E_\mathrm{FS} + E_\mathrm{ex}(\frac{5}{2},\frac{5}{2};0),
\end{equation}
where $E_\mathrm{Mn}$ denotes the energy of the uncoupled Mn adatom and 
$E_\mathrm{FS}$ the energy of the unperturbed Fermi sea. Similarly, the intermediate state has energy 
\begin{equation}
      E^{(0)}_\mathrm{s,out} = 2 E_\mathrm{Mn} + |\epsilon_d| + E_\mathrm{FS} + \xi_k  
        + E_\mathrm{ex}(2,\frac{5}{2};\frac{1}{2})
\end{equation}
for tunneling out and 
\begin{equation}
      E^{(0)}_\mathrm{s,in} = 2 E_\mathrm{Mn} + \epsilon_d + U + E_\mathrm{FS} -
      \xi_k  + E_\mathrm{ex}(2,\frac{5}{2};\frac{1}{2})
\end{equation}
for tunneling in. Here, $-\epsilon_d>0$ is the energy to remove an electron from the filled $d$-shell and $\epsilon_d +U$ the energy to add an electron. We can then compute the perturbative shift of the singlet state as 
\begin{eqnarray}
  \Delta E_s  &=& 2 |V_0|^2 \left\{ \sum_{\xi_k> 0} \frac{1}{[2 E_\mathrm{Mn} + E_\mathrm{FS} + E_\mathrm{ex}(\frac{5}{2},\frac{5}{2};0)] - [2 E_\mathrm{Mn} + |\epsilon_d|+ E_\mathrm{FS} + \xi_k + E_\mathrm{ex}(2,\frac{5}{2};\frac{1}{2})]  } 
   \right.
   \nonumber\\
  && +  \left.  \sum_{\xi_k< 0} \frac{1}{[2 E_\mathrm{Mn} + E_\mathrm{FS} + E_\mathrm{ex}(\frac{5}{2},\frac{5}{2};0)] - [2 E_\mathrm{Mn} + \epsilon_d+ U + E_\mathrm{FS} - \xi_k + E_\mathrm{ex}(2,\frac{5}{2};\frac{1}{2})]} \right\}.
\end{eqnarray}
Note that the two intermediate states $|2,\frac{5}{2}; \frac{1}{2},\pm\frac{1}{2}\rangle$ give the same contributions, each with a factor $1/2$ due to the matrix elements. Note also that the overall factor of two appears, since electrons can tunnel from either Mn adatom of the dimer. We can then simplify
\begin{eqnarray}
  \Delta E_s  &=& - 2 \nu_0 |V_0|^2 \int_0^\infty d\xi  \left\{  \frac{1}{  |\epsilon_d| + \xi + E_\mathrm{ex}(2,\frac{5}{2};\frac{1}{2})  - E_\mathrm{ex}(\frac{5}{2},\frac{5}{2};0)   } 
+   \frac{1}{\epsilon_d + U  + \xi + E_\mathrm{ex}(2,\frac{5}{2};\frac{1}{2}) - E_\mathrm{ex}(\frac{5}{2},\frac{5}{2};0)  } \right\} \quad
\end{eqnarray}
or
\begin{eqnarray}
  \Delta E_s  &=& - 2 \nu_0 |V_0|^2 \int_0^\infty d\xi  \left\{  \frac{1}{  |\epsilon_d| + \xi + \frac{7}{4}J_D   } 
+   \frac{1}{\epsilon_d + U  + \xi + \frac{7}{4}J_D  } \right\} .
\end{eqnarray}
Here, we introduced the density of states $\nu_0$. Assuming the dimer coupling $J_D$ to be small compared to the atomic-physics scales $|\epsilon_d|$ and $U$, we find  
\begin{eqnarray}
  \Delta E_s  &=& \textrm{const} + \frac{7J_D }{4}    2 \nu_0 |V_0|^2 \left\{  \frac{1}{  |\epsilon_d|     } 
+   \frac{1}{\epsilon_d + U   } \right\} ,
\end{eqnarray}
where the constant is a contribution that is independent of the exchange couplings and that cancels out in the singlet-triplet spacing against a similar contribution to the shift of the triplet state. 

Now consider the shift of the triplet state. There are intermediate states with different energies, which have to be incorporated with the appropriate matrix elements. This yields
\begin{eqnarray}
  \Delta E_t  &=& 2 |V_0|^2 \left\{ \sum_{\xi_k> 0} \frac{\frac{7}{15}}{[2 E_\mathrm{Mn} + E_\mathrm{FS} + E_\mathrm{ex}(\frac{5}{2},\frac{5}{2};1)] - [2 E_\mathrm{Mn} + |\epsilon_d|+ E_\mathrm{FS} + \xi_k + E_\mathrm{ex}(2,\frac{5}{2};\frac{1}{2})]  } 
   \right.
   \nonumber\\
  && +   \sum_{\xi_k< 0} \frac{\frac{7}{15}}{[2 E_\mathrm{Mn} + E_\mathrm{FS} + E_\mathrm{ex}(\frac{5}{2},\frac{5}{2};1)] - [2 E_\mathrm{Mn} + \epsilon_d+ U + E_\mathrm{FS} - \xi_k + E_\mathrm{ex}(2,\frac{5}{2};\frac{1}{2})]} 
  \nonumber\\
  && + \sum_{\xi_k> 0} \frac{\frac{8}{15}}{[2 E_\mathrm{Mn} + E_\mathrm{FS} + E_\mathrm{ex}(\frac{5}{2},\frac{5}{2};1)] - [2 E_\mathrm{Mn} + |\epsilon_d|+ E_\mathrm{FS} + \xi_k + E_\mathrm{ex}(2,\frac{5}{2};\frac{3}{2})]  } 
   \nonumber\\
  && +  \left.  \sum_{\xi_k< 0} \frac{\frac{8}{15}}{[2 E_\mathrm{Mn} + E_\mathrm{FS} + E_\mathrm{ex}(\frac{5}{2},\frac{5}{2};1)] - [2 E_\mathrm{Mn} + \epsilon_d+ U + E_\mathrm{FS} - \xi_k + E_\mathrm{ex}(2,\frac{5}{2};\frac{3}{2})]}
  \right\}.
\end{eqnarray}
Using the energies
\begin{eqnarray}
   E_\mathrm{ex}(\frac{5}{2},\frac{5}{2};1) &=& -\frac{31 J_D}{4}   \\
   E_\mathrm{ex}(2,\frac{5}{2};\frac{1}{2}) &=& -7 J_D  \\
   E_\mathrm{ex}(2,\frac{5}{2};\frac{3}{2}) &=& -\frac{11 J_D}{2}  , 
\end{eqnarray}
we find, by the same steps as for the singlet shift, 
\begin{equation}
  \Delta E_t  = \textrm{const} + \left\{ \frac{7}{15} \frac{3J_D }{4} + \frac{8}{15} \frac{9J_D }{4}   \right\}   2 \nu_0 |V_0|^2 \left\{  \frac{1}{  |\epsilon_d|     } 
+   \frac{1}{\epsilon_d + U   } \right\} 
 = \textrm{const} +  \frac{31 J_D}{20}    2 \nu_0 |V_0|^2 \left\{  \frac{1}{  |\epsilon_d|     } +   \frac{1}{\epsilon_d + U   } \right\}.
\end{equation}

Combining results, we obtain the singlet-triplet splitting
\begin{equation}
  \Delta = J_D + \Delta E_t - \Delta E_s
      = J_D\left\{ 1 - \frac{1}{5}    2 \nu_0 |V_0|^2 \left[  \frac{1}{  |\epsilon_d|     } 
     +   \frac{1}{\epsilon_d + U } \right] \right\}.
\end{equation}
Schrieffer \cite{SSchrieffer1967} has derived the $sd$ exchange coupling $J$ between adatom spins (magnitude $S$) and conduction electrons and finds
\begin{equation}
     J = \frac{|V_0|^2}{2S} \left[ \frac{1}{|\epsilon_d|} + \frac{1}{\epsilon_d + U} \right]
\end{equation}
(assuming dominant coupling to a single channel). Thus, we can express the renormalized singlet-triplet splitting as
\begin{equation}
  \Delta = J_D + \Delta E_t - \Delta E_s
      = J_D ( 1 - 2 \nu_0 J ).
\end{equation}
Accounting for the coupling of the adatom to all five conduction electron channels $m$, this result generalizes to
\begin{equation}
  \Delta = J_D ( 1 - 2 \sum_m  \nu_0 J_m ).
\end{equation}
This equation is quoted in the main text. 

\section{Additional experimental data}

\subsection{Adsorption structure of Mn atoms on \mos}

\begin{figure}[h]
  \centering
  \includegraphics[width=0.8\textwidth]{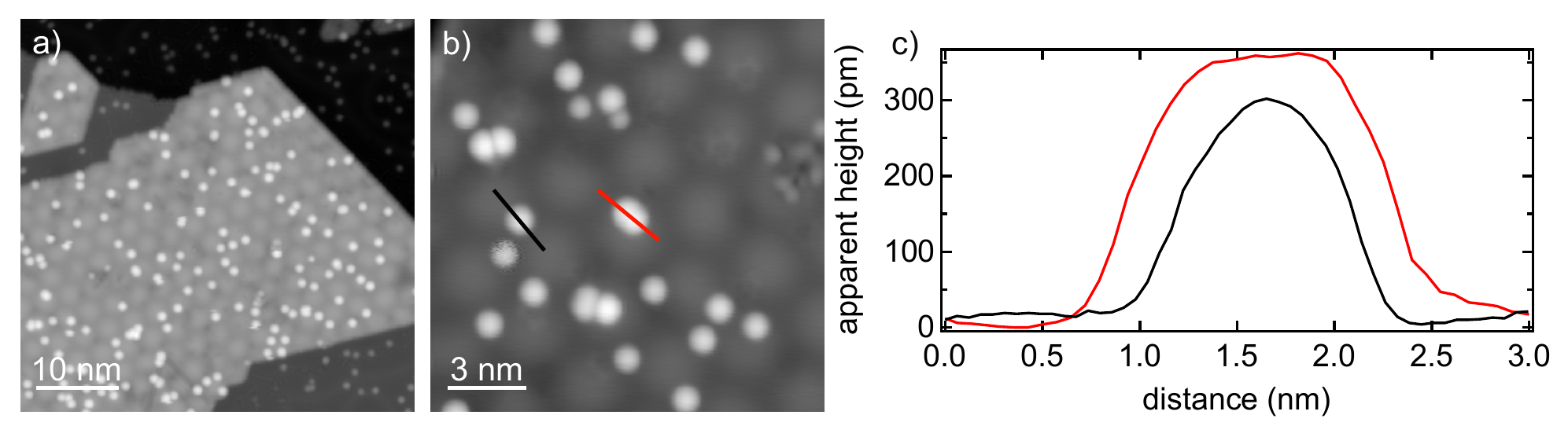}
  \caption{a) Large-scale STM image of a monolayer-island of \mos\ on Au(111) after adsorption of Mn atoms at low temperature. Recorded at 1 V and 100 pA. b) Close-up view showing individual atoms as round protrusions and some elongated structures most probably being Mn dimers. Some point defects can be observed in the \mos\ layer. Recorded at 100 mV and 20 pA. c) Height profiles along the black and red lines shown in b. 
		}
  \label{MNF1}
\end{figure}

Figure \ref{MNF1}a shows an overview topography image of a  monolayer-island of \mos\ decorated with a large number of Mn atoms. A close-up view confirms that the individual atoms appear as round protrusions throughout a bias voltage range of -1 to 1 V (Fig.\ \ref{MNF1}b). Owing to the convolution with the tip shape, the atoms appear with a large width ($\sim$ 0.9 nm), impeding the determination of the exact adsorption site on the atomic lattice constant of \mos.
The similarity of apparent heights and spectroscopic signatures suggests that all atoms adsorb in equivalent lattice sites. This is in agreement the observation of unique adsorption sites of Fe on \mos\ \cite{STrishin2021}. DFT calculations further suggest hollow sites to be the energetically most favorable positions \cite{SChen2017,SWang2014}. Occasionally, we find elongated protrusions (see also lineprofiles in Fig.\ \ref{MNF1}c), which we ascribe to dimers.

\clearpage

\subsection{Manipulation of Mn atoms}

\begin{figure}[h]
  \centering
  \includegraphics[width=0.8\textwidth]{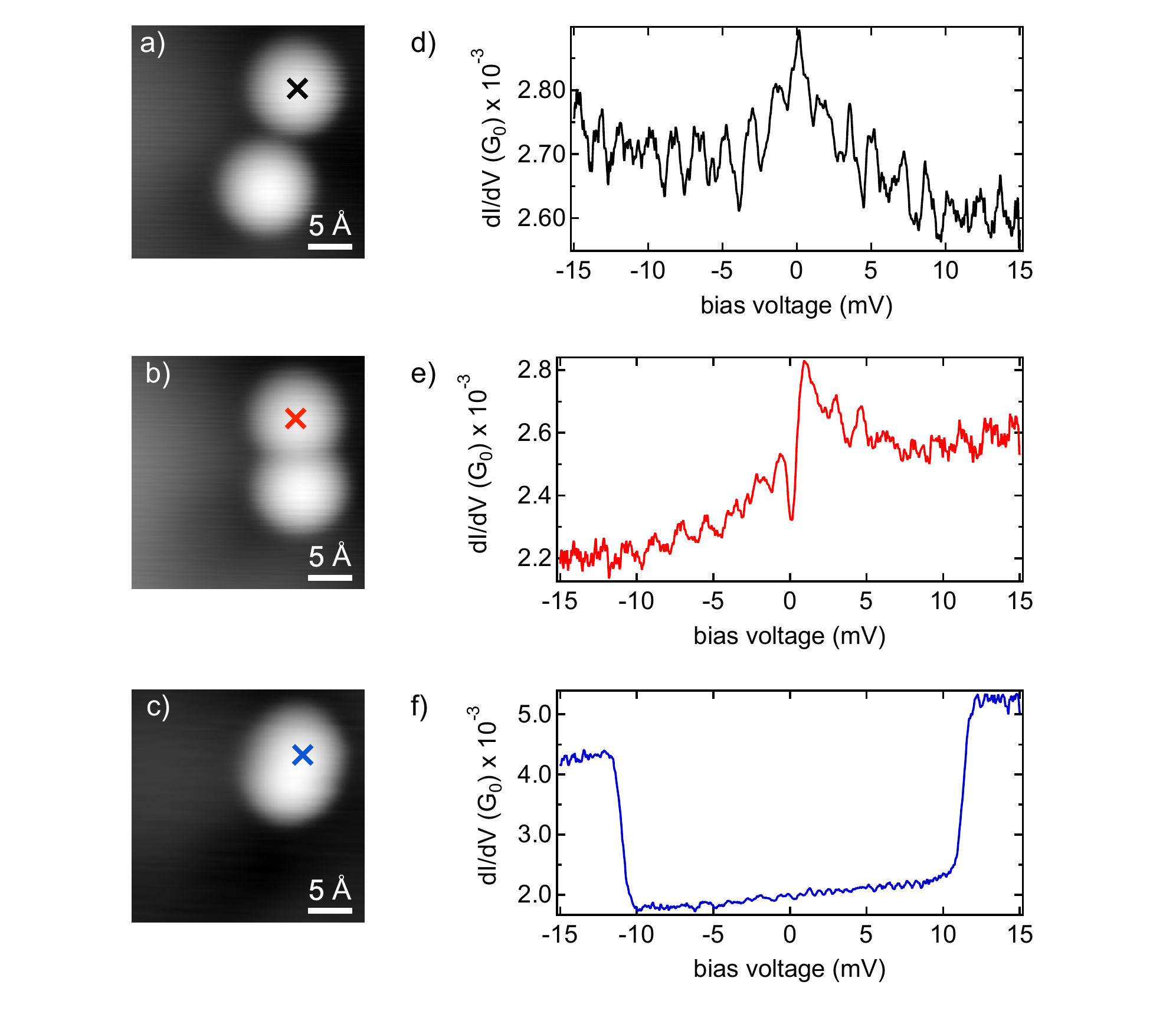}
  \caption{Manipulation of two Mn atoms into dimer structures. a-c) STM topographies of the same atoms before and after successive manipulation events. The atom at the bottom of figure (a) was pushed closer towards the other upper atom, as seen in (b). Here the atoms are still distinguishable. In (c) the lower atom was pushed even closer to the upper atom, resulting in a dimer. d-f) \didv spectra performed on the upper atom in (a), (b) and (c) respectively. The topographies were recorded at 100 mV and 20 pA, the setpoint of the recorded spectra was 15 mV and 3 nA (f) and 10 mV and 3 nA (g). }
  \label{MN7}
\end{figure}

We mainly investigated Mn dimers statistically distributed over the surface. In rare cases, we were able to manipulate the Mn atoms in a controlled manner. Fig.\ \ref{MN7} shows an example of consecutive manipulation events and  the \didv spectra recorded on the obtained structures. In Fig.\ \ref{MN7}a two Mn atoms are separated at sufficiently far distance such that they exhibit a Kondo resonance (spectrum shown in \ref{MN7}d). At closer distance (b), the Kondo resonance is split (Fig.\ \ref{MN7}e). When the atoms are pushed into adjacent lattice sites as in Fig.\ \ref{MN7}c, the singlet-triplet excitation is observed (Fig.\ \ref{MN7}f).

\begin{figure}[h]
  \centering
  \includegraphics[width=0.8\textwidth]{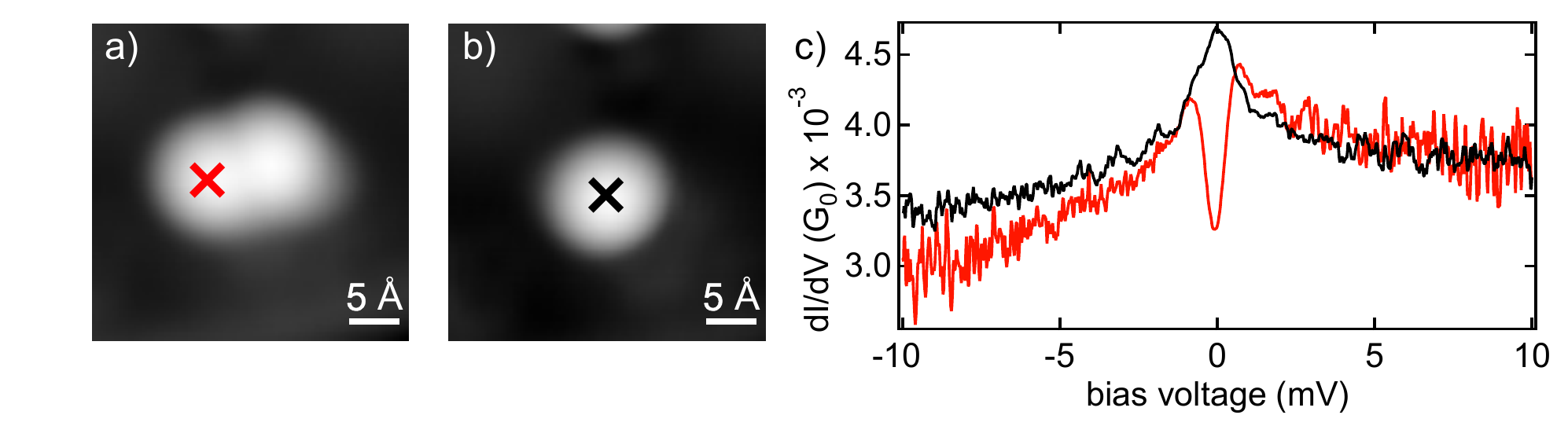}
  \caption{Disassembly of a Mn dimer. a,b) STM topographies of a Mn dimer before and after the removal of one atom. Here the right atom in (a) was removed, leading to a single Mn atom as shown in (b). c) \didv spectra performed on the left atom in (a) and on the same (remaining) atom (b) respectively. The topographies were recorded at 100 mV and 20 pA, the setpoint of the recorded spectra was 10 mV and 3 nA (g). }
  \label{MN8}
\end{figure}

Fig.\ \ref{MN8}a shows one dimer where two Mn atoms are two lattice sites apart. The Kondo resonance is split (red line in Fig.\ \ref{MN8}c). Removing one of the atoms leads to an unperturbed Kondo resonance (green line in Fig.\ \ref{MN7}c).

\begin{figure}[h]
  \centering
  \includegraphics[width=0.8\textwidth]{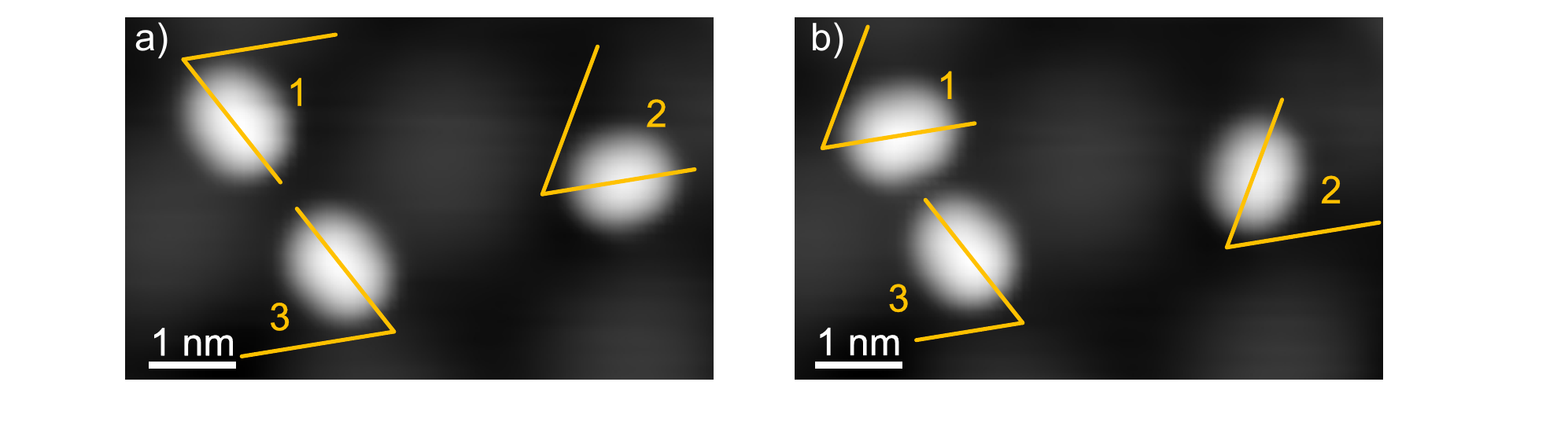}
 \caption{ Rotation of Mn dimers. a), b) STM topographies of single Mn dimers before (a) and after (b) applying a high bias voltage. In (a) the dimers 1 and 3 show the same orientation,  whereas dimer 2 is rotated by roughly  120$^\circ$ with respect to 1 and 3. After a bias voltage of 1.5 V was applied to the dimers in (a), dimer 1 and 2 appear rotated by 120$^\circ$. The topographies were recorded at 100 mV and 20 pA. }
  \label{MN13}
\end{figure}

An unambiguous assignment of the adsorption sites of the Mn atoms within the dimer structures is challenging as the Mn atoms appear very large and cannot be separately resolved. Analyzing the orientation of the dimers on the surface, we observed only three orientations, suggesting the registry with the threefold atomic lattice structure of \mos. While attempting to remove one of the Mn atoms from the densely-packed dimer structures by a voltage pulse, we often observed effectively a rotation of the dimers. Also the resulting dimers follow the main axes (Fig. \ref{MN13}).

\subsection{RKKY coupled dimers in different moir\'e sites}

\begin{figure}[h]
\includegraphics[width=0.5\columnwidth]{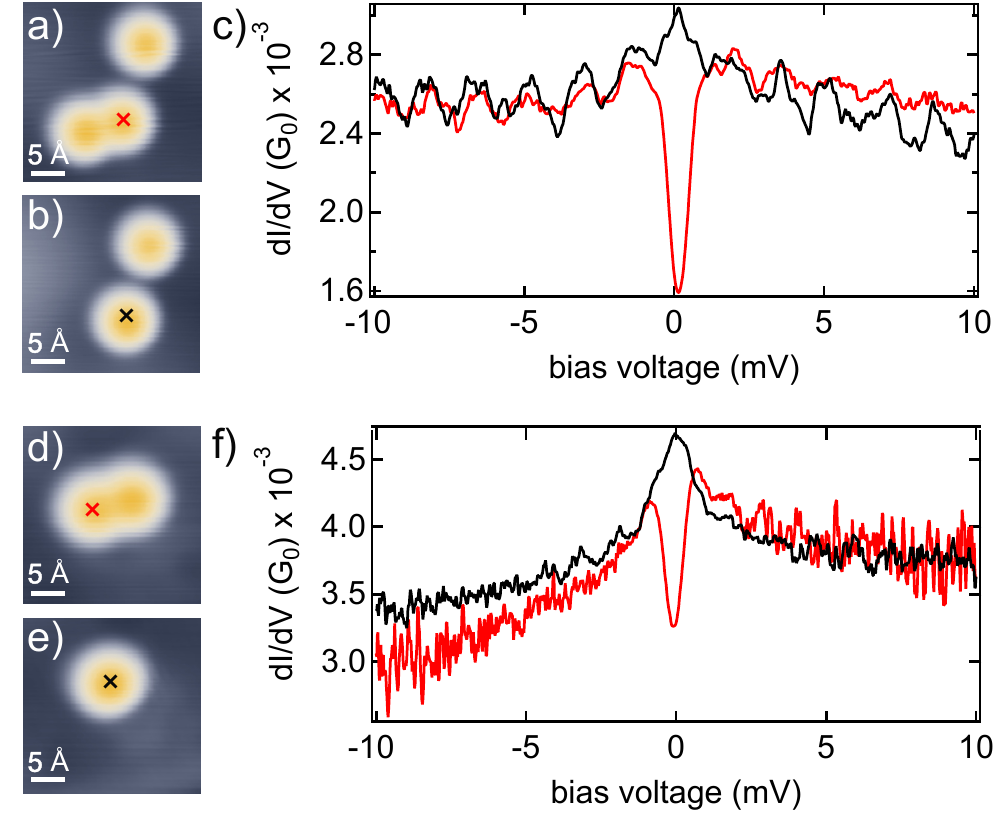}
\caption{Moir\'e effect on RKKY-coupled Mn dimers. a), d) STM topographies of Mn dimers. Whereas in (a) the dimer is adsorbed close to the moir\'e maximum, in (d) the dimer is adsorbed in the moir\'e valley. c), f) \didv spectra performed at the crosses in (a) and (d). b),e) show the same scan frame, after one atom has been removed from the dimer. The black spectra in (c) and (f) show the spectra of the respective monomer. The topographies were recorded at 100 mV and 20 pA, the setpoint of the recorded spectra was 15 mV and 3 nA (c) and 10 mV and 3 nA (f). }
 \label{figS4}
\end{figure}

In the main text, we showed the variation of singlet-triplet excitations along the moir\'e superstructure. To probe whether RKKY-coupled Mn dimers are equally affected by the moir\'e structure, we investigate Mn dimers with a spacing of two substrate lattice sites (Fig. \ \ref{figS4}). As described in the main text, substrate-mediated interactions lead to small excitation gaps around the Fermi level on top of the Kondo resonance (red lines in Fig.\ \ref{figS4}c,f). Various dimers in different moir\'e sites display similar gap sizes while the height of the Kondo resonance varies. The same height modulation of the Kondo resonance is found on the isolated atoms in the same adsorption sites. This is shown by spectra taken on the same atoms after the neighbor has been removed by STM manipulation (black lines in Fig.\ \ref{figS4}c,f). Hence, once Kondo correlations of the individual atoms dominate the spectra and the coupling enters through a small perturbation, we hardly observe any moir\'e induced modulations in the coupling.


%

\end{document}